# Three dimensional valency mapping in $CeO_{2-x}$ nanocrystals

B. Goris[†], S. Turner[†], S. Bals[*] and G. Van Tendeloo

EMAT, University of Antwerp, Groenenborgerlaan 171, B-2020 Antwerp, Belgium

* sara.bals@uantwerpen.be

**Using electron tomography in combination with spatially resolved electron energy-loss spectroscopy at high energy resolution, we are able to map the valency of the Ce ions in $CeO_{2-x}$ nanocrystals. Our three-dimensional results show a clear facet-dependent reduction shell at the surface of ceria nanoparticles; {111} surface facets show a low surface reduction, whereas at {001} surface facets the cerium ions are more likely to be reduced over a larger surface shell. The novelty of this generic tomographic technique is that it allows a full three dimensional datacube to be reconstructed, containing a full electron energy-loss spectrum in each voxel. The ability to extract an electron energy-loss spectrum in each point of a reconstructed datacube, enables the three-dimensional investigation of a plethora of material-specific physical properties such as valency, chemical composition, oxygen coordination or bond lengths. These experiments will trigger the synthesis of nanomaterials with improved properties and the design of nanostructures with novel functionalities.**

Cerium oxide (ceria) nanoparticles have an enormous potential in a variety of materials science applications. For example, the addition of ceria nanoparticles to diesel is known to drastically reduce soot in exhaust streams[1]. The most advanced applications are the use as a three-way catalyst to remove unwanted by-products like $NO_x$, CO, and other unreacted hydrocarbons from exhaust fumes and use as a chemo-mechanical polishing agent in the microelectronics industry[2]. Doped cerium dioxide materials are also attracting ever-increasing interest, with applications in the field of electrolytes for solid oxide fuel cells (SOFCs) and as base catalyst for hydrogen production by the process of chemical looping[3,4].

Any successful implementation of ceria nanoparticles in current and future applications strongly depends on a thorough understanding of the connection between the physical properties and the local three dimensional (3D) structure and composition of the material. It is known that the catalytic activity of ceria is determined by its flexible reduction and oxidation behaviour. This is connected to the possibility of switching between $Ce^{4+}$ and $Ce^{3+}$ oxidation states and the corresponding ability to release and take up oxygen at the surface of the nanoparticles through the formation of oxygen vacancies. Surprisingly, the exact connection between the 3D surface structure and the oxygen storage capability is still poorly understood. Precise measurement of the oxidation state of the near-surface cations in catalytic nanoparticles can therefore be considered a missing link in the detailed understanding of their behaviour. Until now, such studies could not be performed due to a lack of the required 3D analytical techniques at such a local scale.

Electron energy-loss spectroscopy (EELS) can provide information about the valency of cations, through the shape and positions of ionisation edges within the EELS spectrum. Recent advances in instrumentation such as the implementation of electron monochromators, allows EELS edges to be acquired at energy resolutions close to that of synchrotron radiation. This means that, on top of e.g. valency, other information like oxygen coordination and bond elongation can be extracted from core-loss spectra, whereas plasmonics, interband transitions and band gaps can be studied in the low-loss regime. The fact that this energy resolution is available in instruments which can form atom-sized resolution probes means that spectroscopic information can be mapped out at atomic resolution[5-7].

Recently, valency changes in cerium nanoparticles were investigated by aberration corrected scanning transmission electron microscopy (STEM) in combination with spatially resolved EELS[8]. In this manner, a change from $Ce^{4+}$ to $Ce^{3+}$ could be observed and the extent of the reduced shells was investigated as a function of particle size. Although this study is already at the forefront of electron microscopy characterisation, one should never forget that a single valency map obtained by electron microscopy only corresponds to a two dimensional (2D) projection of a 3D object. As the catalytic behaviour is known to depend on the specific type of crystallographic surface facets, it is of crucial importance to determine the exact relation between the nanocrystal morphology, surface structure and the cation oxidation states. In their previous work the authors postulated on the shell thickness as a function of the type of surface facet, but unambiguous conclusions could impossibly be drawn from 2D projection data only.

3D electron microscopy, or so-called "electron tomography" is a technique that yields a 3D reconstruction of a (nano)material based on a tilt series of its 2D projection images. For crystalline specimens, high angle annular dark field (HAADF) STEM projection images are routinely used, resulting in 3D reconstructions of the morphology as well as the inner structure of a broad range of materials[9,10]. Most results are at the nanometer scale, but recently also 3D reconstructions with atomic resolution have been obtained[11-16]. In addition to a 3D structural characterisation, analytical information in 3D can also be obtained by expanding Energy dispersive X-ray spectroscopy (EDX) and EELS from 2 to 3D[17-21]. An exciting example of this type of work is the 3D visualisation of plasmon modes of a Ag nanocube[22].

Although not straightforward, a 3D visualisation of the $Ce^{3+}/Ce^{4+}$ distribution in ceria nanoparticles could in principle be obtained using 2D maps valency maps as an input for the reconstruction. However, using such an approach, part of the information that is available from the tilt series of EELS cubes is omitted. Furthermore, when extracting a 2D valency map from the EELS data cube, systematic errors are likely to occur. During the tomographic reconstructions, these errors will be accumulated, hampering a reliable 3D quantification. Another more generic and comprehensive possibility is to use complete EELS data cubes as an input for tomographic reconstruction, leading to a 4 dimensional (4D) data set from which an energy loss spectrum can be extracted for each reconstructed voxel. Such experiments are challenging since they require the combination of advanced spectroscopic and tomographic techniques in a single experiment. Also computationally such a reconstruction is clearly very demanding, but the development of reconstruction techniques based on graphics processing units (GPU) makes this approach feasible[23]. The latter methodology however, prevents an accumulation of errors that can be introduced during the extraction of the 2D projection maps and therefore, 3D quantitative results can be obtained in a more straightforward and reliable approach. The 3D $Ce^{3+}$ vs. $Ce^{4+}$ distribution can be determined by fitting the reconstructed EELS spectrum in each voxel to known references for $Ce^{3+}$ and $Ce^{4+}$ in a linear combination. These novel experiments will lead to unique insights into the relationship between the amount of cerium atoms that are in a reduced state and the thickness of the reduction shell versus the different crystallographic surface facets. In this work, we will study sharply faceted ceria nanoparticles from gas-phase synthesis, in order to investigate the connection between the most common surface facets and the degree of cerium reduction measured using our novel technique. Although we apply our approach to the investigation of $CeO_{2-x}$ nanoparticles, it must be noted that our generic methodology is applicable to a wide variety of nanostructures and will undoubtedly lead to a wealth of new information.

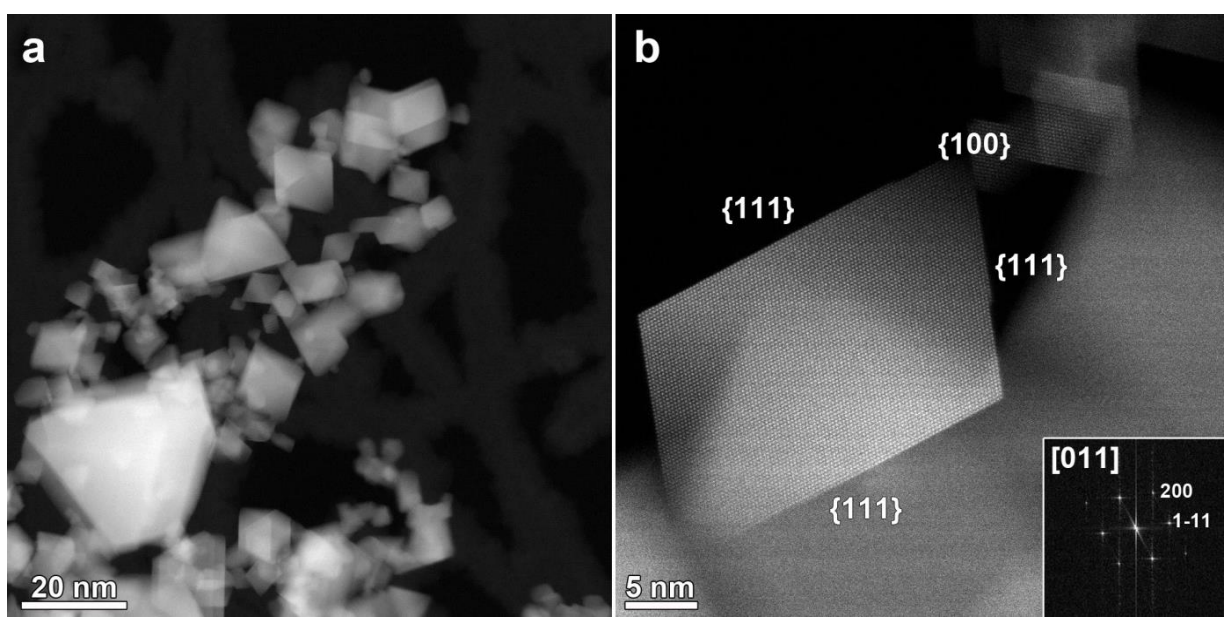

**Figure 1: HAADF-STEM projection images of ceria nanoparticles.** (a) HAADF-STEM overview image of the ceria nanopowder used in this study. The ceria nanoparticles are sharply faceted. (b) High resolution HAADF-STEM image of a single ceria nanoparticle, imaged along the [011] zone axis orientation, as evidenced by the inset Fourier Transform pattern. The main surface facets are {111}-type.

An overview image of the sample, acquired using HAADF-STEM is presented in **Figure 1a**. It can be seen that the ceria nanoparticles are agglomerated, with sizes ranging from ~ 10 to 50 nm in diameter. The high resolution HAADF-STEM image in **Figure 1b** shows that the nanoparticles are highly faceted, with the main terminating planes being {111}-type. The image demonstrates that the nanoparticle is single-phase, no core-shell type structural changes are present at the nanoparticle surface. If a valency changes can be registered at the nanoparticle surface, it must then be related to the presence of oxygen vacancies, and not to a structural change. It is clear that from a single 2D projection image, it is impossible to determine the exact 3D morphology and degree of truncation of the nanoparticle. Therefore, tilt series of projection images were acquired for two different ceria nanoparticles. More details on the experimental conditions are provided in the methods section. A 2D projection image from the tilt series is presented for each nanoparticle in **Figures 2a** and **b**. Additional projection images extracted from both tilt series are provided in **Supplementary Figure 1**. After alignment of the tilt series, 3D reconstructions were computed as explained in the methods section. Visualisations of the 3D reconstructions are presented in **Figure 2c** and **d**. These 3D results reveal that the first particle yields a near-perfect octahedral morphology consisting of eight {111} facets, whereas the second particle shows a truncation along the [100] direction as indicated in **Figure 2d**. For both particles, an overlap with another particle at the left can be observed.

Since ceria {111} surfaces are known to yield lower catalytic activity in comparison to {001} surfaces[24], it is of great interest to compare the extent of the reduced shell for both particle morphologies. Therefore, EELS datacubes were acquired, from which spectra (**Figures 2f**) were extracted at the positions indicated by the green and red squares in **Figures 2b**. By fitting the acquired EELS data to reference spectra (**Figure 2e**), 2D oxidation maps corresponding to $Ce^{3+}$ and $Ce^{4+}$ were obtained (**Figure 2g-j**). This procedure is explained in more detail in the methods section and in previous work[8]. From **Figure 2h** and **j**, a reduced shell containing $Ce^{3+}$ ions at the surface of the nanoparticle is obvious. However, a quantitative interpretation of the shell thickness for the different surface facets is not possible, as the facets are not imaged edge-on. Even when the particles would be imaged edge-on at high resolution, it would still be unreliable to measure the thickness of the $Ce^{3+}$ layer at small facets that are present in the morphology because of thickness effects that are inherent to the projection principle. Furthermore, the overlap of different particles as observed on the left of the projection images in **Figures 2a** and **b** prohibits a clear understanding of the reduction effect at the interface between both particles.

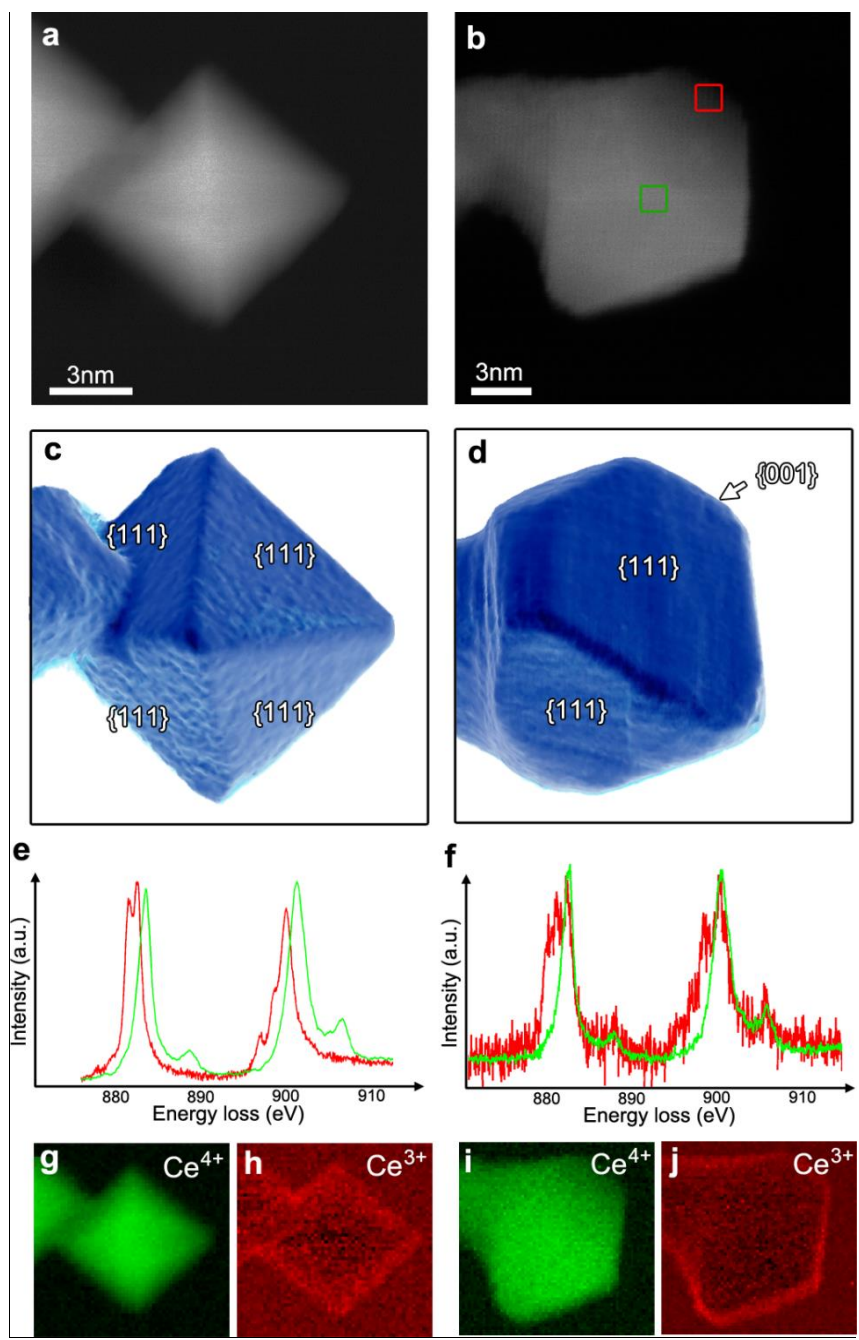

**Figure 2: Combination of 2D EELS and HAADF-STEM tomography results.** (a and b) HAADF-STEM projection images of ceria nanoparticles with an octahedral and a truncated octahedral morphology. (c and d) 3D visualisations of the HAADF-STEM tomographic reconstructions. (e) Reference EELS spectra for $Ce^{3+}$ (red) and $Ce^{4+}$ (green) (f) Spectra from the EELS data cube of the truncated octahedral particle (positions indicated in b). Based on the reference spectra for $Ce^{3+}$ and $Ce^{4+}$, 2D

projection maps can be obtained containing information about the valency of both nanoparticles. (g-j)

Obviously, a single 2D map of the Ce oxidation state is insufficient when trying to connect the role of oxygen vacancies to the nanoparticle morphology. A clear understanding of this role consequently relies upon the expansion of these experiments into 3 dimensions. Therefore, a tilt series of EELS data cubes was acquired under conditions that are provided in the Methods section. In order to be suitable for 3D reconstruction, the projection requirement states that the 2D images should yield an image intensity that is a monotonic function of sample thickness. In this study, the small size of the investigated nanoparticles (±10nm) guarantees that multiple scattering effects and beam spreading are minimal and therefore the projection requirement is fulfilled. In order to compensate for spatial drift during the acquisition of each map, an affine transformation was applied on the EELS data prior to the electron tomography reconstruction. This procedure is explained in the Methods section and illustrated in **Supplementary Figure 2**.

The final reconstructed result corresponds to a 4D data cube in which each (3D) voxel contains a complete EELS spectrum. It is important to point out that no filtering, noise suppression or any other data processing was applied to the EELS data prior to, or during tomographic reconstruction. The 4D data reconstruction is illustrated in **Figure 3**, where the morphology of the truncated ceria nanoparticle is presented along with a spectrum that is extracted from the 3D region indicated in **Figure 3a**. From **Figure 3b**, it is obvious that small differences in the spectra can be observed for voxels that are located in the inner part of the reconstruction and voxels that are located at the boundary. These differences directly correspond to a $Ce^{4+}/Ce^{3+}$ surface reduction.

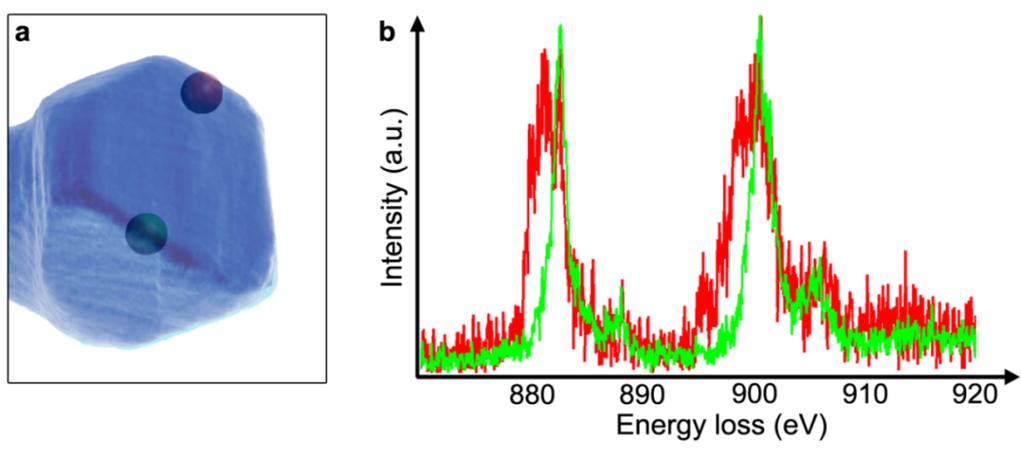

**Figure 3: Extracted EELS spectra from a 3D volume.** (a) Three dimensional visualisation of the morphology of the reconstructed nanoparticle. At two different positions (indicated in a), a 3*3*3 voxel averaged EELS spectrum is extracted and presented in (b).

Similar to the procedure that was performed to obtain 2D oxidation maps from the 3D data cube, we are now able to fit the reference spectra to each 3D voxel in the reconstructed volume. As a result, the 3D spatial distribution of $Ce^{3+}$ and $Ce^{4+}$ can be obtained for both samples as presented in **Figure 4a** and **d**. **Figures 4b** and **e** show 3D visualisations of the regions corresponding to $Ce^{3+}$ (red) and $Ce^{4+}$ (green). In order to quantify the shell thickness, slices through the reconstruction can now be extracted from the data. Relevant examples are presented in **Figures 4c** and **f**. From these figures, a uniform shell thickness for the {111} facets is observed. This is obvious for the octahedral particle, which demonstrates an even $Ce^{3+}$ signal on all facets. The tomographic nature of the technique used in this work also provides information on internal valency changes within the material. In this manner, the boundary between the two nanoparticles in Figure 4.a is demonstrated to remain fully oxidised. It must be noted that this conclusion could not be drawn from the projected valency maps in Figure 2. For the truncated nanoparticle, the shell is observed to be thicker along a

{001} surface plane in comparison to the {111} planes, as indicated in **Figure 4f**. Similar observations could be made for an additional truncated particle that was investigated. These results are presented in **Supplementary Figure 3**.

Normalized references were used to provide the spectral weights for the $Ce^{3+}$ and $Ce^{4+}$ components in each of the reconstructed voxels. This approach has the benefit of delivering quantitative data for the cerium oxidation state. This means that the slices presented in **Figure 4c** and **f** not only show where the surface of the nanoparticles is reduced, but moreover to what extent. The quantitative data shown in **Figure 4c** evidences that at a {111} surface plane approximately 20-30 percent of the ceria ions are reduced. The reduction shell has an approximate thickness of 0.8±0.2 nm. It is fascinating to see that at corners the surface reduction can be elevated, up to a maximum value at the bottom corner of the octahedron of 54 percent Ce reduction (arrows). In the case of the truncated octahedron (**Figure 4f**), the {001} surface facet shows a higher degree of surface reduction (±50 percent Ce ions in a reduced state) over a thicker measured shell (1.4±0.2 nm). Once again, the corners formed by the {111} surface facets show a tendency for a higher degree of Ce reduction (arrows).

The different extent of surface reduction for {001} and {111} facets can be understood based on the atomic arrangement of ceria in a fluorite structure, in which the {111} facets are the most closely packed. As a consequence, oxygen atoms are more heavily hindered during the reduction or oxidation of the surface layers, resulting in a thinner $Ce^{3+}$ layer. These results provide a direct explanation for the observed lower catalytic activity of the {111} surfaces with respect to the {001} surfaces.

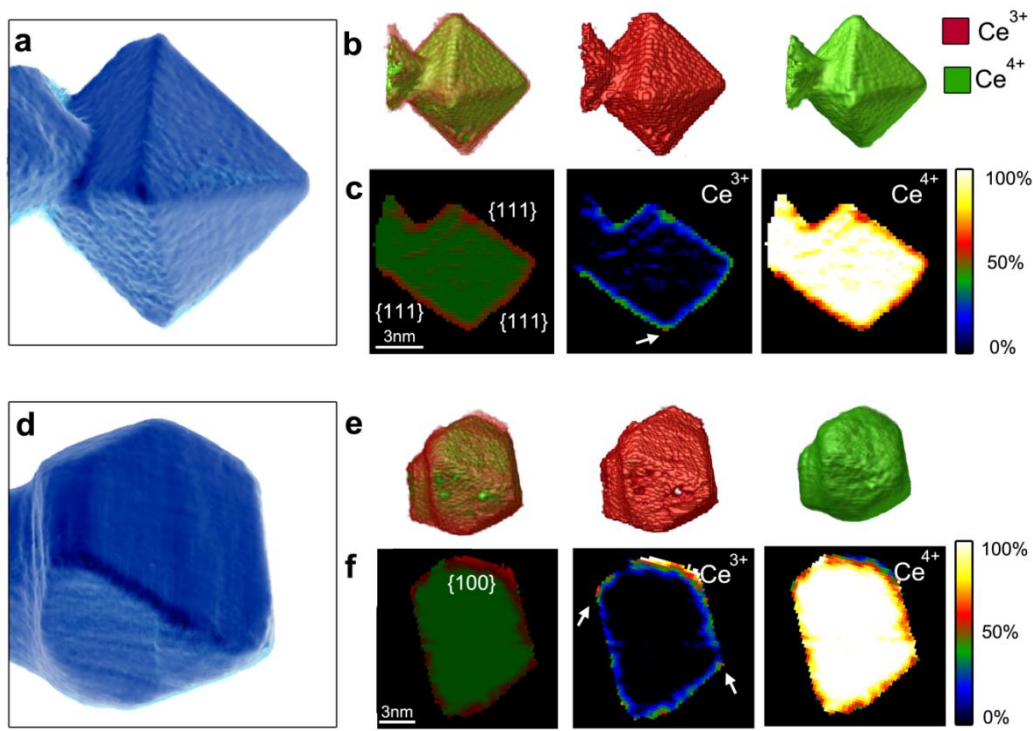

**Figure 4: 3D valency measurements**. (a and d) HAADF-STEM reconstructions of near-perfect and truncated octahedral ceria nanoparticle. (b and e) The corresponding 3D visualisations and slices through the 3D reconstructions showing the valency results for $Ce^{3+}$ and $Ce^{4+}$, indicating a thicker $Ce^{3+}$ layer with the presence of more oxygen vacancies at the {001} truncation. (c and f) Slices through the $Ce^{3+}$ and $Ce^{4+}$ volumes, yielding a quantitative distribution of the reduced Ce ions.

In conclusion, we have mapped the valency of the Ce ions in $CeO_{2-x}$ in all three spatial dimensions using electron tomography combined with spatially resolved electron energy-loss spectroscopy at high energy resolution. These unique experiments reveal a clear facet-dependent reduction shell at the surface of ceria nanoparticles, which is invisible to modern high resolution transmission electron microscopy structural imaging techniques. The main

{111} surface facets show a low surface reduction, whereas at {001} surface facets the cerium ions are more likely to be reduced over a larger surface shell.

The generic and innovative approach proposed in this study holds enormous potential, as it can be applied to a wide range of material science problems. The methodology opens the door to measurement of material properties such as valency, chemical composition, oxygen coordination and bond lengths in nanomaterials. The unique insights that one is able to obtain in this manner will trigger the synthesis of nanomaterials with improved properties and the design of nanostructures with novel functionalities.

**Methods:**

Ceria nanomaterial:

The ceria particles used for this study are commercial nanograin®, gas-phase produced nano $CeO_{2-x}$ particles from Umicore NV/SA, Belgium. The material was prepared for TEM by dispersing the powder in ethanol and dropping the dispersion onto a holey carbon grid. $CeO_2$ and $CeO_3$ EELS references were purchased from Sigma Aldrich and prepared in a similar way.

STEM-EELS imaging:

All data was acquired using an aberration corrected Titan 60-300 microscope operated at an acceleration voltage of 120 kV. EELS data cubes were acquired at tilt angles ranging from -70º to +65º or -60º to +65º with a tilt increment of 5º. The pixel size equals 2 and 2.3 angstrom respectively, which is smaller than the interatomic distances. The individual spectrum images were 66x66 pixels, taking a dwell time of 0.08s per pixel. The energy resolution provided by the electron monochromator, as measured from the full-width half maximum of acquired zero-loss peaks was 0.2 eV, the dispersion of the spectrometer was set to 0.05 eV and 2048 channels are used to cover an energy range from 845eV to 947eV, being the entire Ce $M_{4,5}$ edge. The convergence semi-angle α used for the experiments was 18 mrad, the HAADF-STEM collection angles and EELS collection semi-angle β was ~60 mrad in order to guarantee incoherent imaging.

Tomography reconstruction:

Prior to their use as an input for tomographic reconstruction, the EELS data cubes have to be corrected for the drift present during the acquisition. This is obtained by comparing an original HAADF-STEM image acquired prior to the collection of the EELS data with a HAADF-STEM image acquired simultaneously with the EELS data. The transformation (linear drift) between both is measured using a least square minimisation and is used to compensate the drift in each energy channel. Using a manual alignment, the HAADF-STEM tilt series is aligned and the alignment parameters are applied to the tilt series of the different energy channels. A reconstruction is calculated using the ASTRA toolbox[23,25], resulting in a separate reconstruction for each energy level. During the reconstruction, the HAADF-STEM result is used as a mask in order to improve the final result.

Determination of $Ce^{3+}$ or $Ce^{4+}$

Once a reconstruction is calculated for each energy level, a 4D data cube is obtained from which an EELS spectrum can be extracted for each individual 3D voxel. These reconstructed spectra are fitted to reference spectra of $Ce^{3+}$ and $Ce^{4+}$, obtained from $CeF_3$ and bulk $CeO_2$ respectively, in a linear combination using a mean least squares fit, in order to determine the presence and strength of each component in every reconstructed voxel. The shell thickness was determined using a threshold value of 5% $Ce^{3+}$.

**Acknowledgements:**

This work was supported by funding from the European Research Council under the Seventh Framework Program (FP7), ERC grant N°246791 – COUNTATOMS and ERC grant N°335078 – COLOURATOMS. The authors gratefully acknowledge the fund for scientific research Flanders (FWO-Vlaanderen). We are grateful to K.J. Batenburg and J. Sijbers for providing the ASTRA toolbox for tomography and useful discussions. Umicore NV/SA (Belgium) is gratefully acknowledged for the provision of the nano ceria samples. S. Put and Y. Strauven are thanked for valuable discussions relating to the material. We thank J. Verbeeck for useful discussions.